\documentclass{aa}
\usepackage{graphicx}
\usepackage{txfonts}
\usepackage{epsfig}
\usepackage{times}
\def\ltsima{$\; \buildrel < \over \sim \;$}
\def\gtsima{$\; \buildrel > \over \sim \;$}
\def\simlt{\lower.5ex\hbox{\ltsima}}
\def\simgt{\lower.5ex\hbox{\gtsima}}

\begin{document}

   \title{Extreme Warm Absorber variability in the Seyfert Galaxy Mrk~704}

   \author{G. Matt \inst{1},  S. Bianchi\inst{1,2}, 
M. Guainazzi\inst{3}, A.L. Longinotti\inst{4}, M. Dadina\inst{5}, 
V. Karas\inst{6}, G. Malaguti\inst{5}, G. Miniutti\inst{7}, 
P.O. Petrucci\inst{8}, E. Piconcelli\inst{9}, G. Ponti\inst{10}
}

   \offprints{G. Matt, \email{matt@fis.uniroma3.it} }

   \institute {$^1$Dipartimento di Fisica, Universit\`a degli Studi Roma Tre, 
via della Vasca Navale 84, I-00146 Roma, Italy \\
$^2$ INAF - Osservatorio Astronomico di Brera, Via E. Bianchi 46, I-23907, Merate, Italy \\
$^3$ European Space Astronomy Center of ESA, Apartado 50727, 28080 Madrid, Spain \\
$^4$ M.I.T. Kavli Institute for Astrophysics and Space Research, Cambridge, USA \\
$^5$ INAF-IASF Bologna, via Gobetti 101, I-40129, Bologna, Italy\\
$^6$ Astronomical Institute, Academy of Sciences, Bocni II 1401, CZ-14131 Prague, Czech Republic \\
$^7$ Centro de Astrobiología (CSIC-INTA) LAEFF, PO Box 78, Villanueva de la Ca\~nada, Madrid 28691, Spain\\
$^8$ UJF-Grenoble 1 / CNRS-INSU, Institut de Planetologie et
            d'Astrophysique de Grenoble (IPAG) UMR 5274, Grenoble, F-38041, France \\
$^9$ INAF - Osservatorio Astronomico di Roma, via Frascati 33, I-00040, Monteporzio Catone, Roma, Italy \\
$^{10}$ School of Physics and Astronomy, University of Southampton, Highfield, Southampton SO17 1BJ, UK
}

   \date{Received / Accepted }

   \abstract
{In about half of Seyfert galaxies, the X-ray emission is absorbed by an optically thin, ionized
medium, the so-called ``Warm Absorber'', whose origin and location is still a matter of debate. }
{The aims of this paper is to put more constraints on the warm absorber by studying its variability.}
{We analyzed the X-ray spectra of a Seyfert 1 galaxy, Mrk~704, which was observed twice,
 three years apart, by XMM-$Newton$.}
{The spectra were well fitted with a two zones absorber, possibly covering only partially the source. 
The parameters of the absorbing matter - column density, ionization state, covering factor - 
changed significantly between the two observations.  Possible explanations for the more ionized absorber
are a torus wind (the source is a polar scattering one) or, in the partial covering scenario, 
an accretion disk wind. The less ionized absorber may be composed of orbiting clouds in the 
surroundings of the nucleus, similarly to what already found in other sources, most notably NGC~1365. 
}{}

\authorrunning{G. Matt et al. }
\titlerunning{Extreme Warm Absorber variability in the Seyfert Galaxy Mrk~704}

   \maketitle
%

\section{Introduction}

Absorption from ionized matter in the X-ray spectrum  
of AGN (the so-called warm absorber) was discovered many years ago (Halpern 1984).
Since then, many advances in its understanding have been made, and we know now that
it is present in about half of Seyfert galaxies (e.g. Reynolds 1997) and that the
matter is photoionized. Warm absorbers are also known to vary, and indeed the first discovered absorber
was variable (Halpern 1984). The location of the warm absorber (or, indeed, of the absorbers,
as more than one ionizing zone is often found) is, however, largely uncertain.
There is some evidence for its origin as a wind from the dusty torus envisaged
in unification models for Seyfert galaxies (Blustin et al. 2005), but cases in which 
an origin from the disk seems to be preferred do also exist (Krongold et al. 2007).

Mrk~704 is a local ($z$=0.029234) Seyfert 1.2 galaxy (Veron-Cetty \& Veron 2010),
bright enough in X-rays to be detected by $Swift$/BAT (Ajello et al. 2008). 
In this paper we
report on extreme warm absorber variability, on yearly time scales, 
revealed by two XMM-$Newton$ observations,
and possible variability on monthly time scales from short Swift/XRT observations.

The paper is organized as follows: in Section~2 we report on the XMM-$Newton$ 
observations and data reduction, while the relative data analysis are discussed
in Section~3. Section~4 presents the analysis of the $Swift$ and $ASCA$
observations, while the results are summarized and discussed in Section~5.

\begin{figure*}
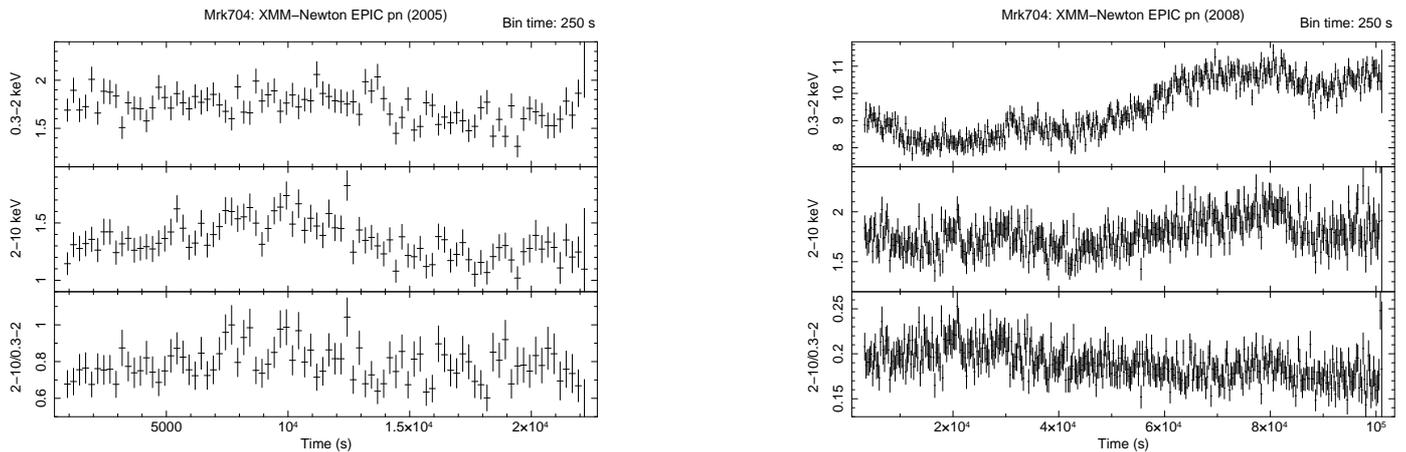

\epsfig{file=lc_obs1.ps,width=6cm,angle=-90}
\hfill
\epsfig{file=lc_obs2.ps,width=6cm,angle=-90}
  \caption{The 0.3-2 keV (upper panel), 2-10 keV (middle panel) and the hard-to-soft ratio 
(lower panel) light curves, for the first (left panels) and second (right panels) 
XMM-$Newton$ observations.}
  \label{lc_obs}
\end{figure*}

\section{XMM-$Newton$: Observations and data reduction}

XMM-$Newton$ observed Mrk~704 twice, on 2005-10-21 (\textsc{obsid}: 0300240101) and on
2008-11-02 (\textsc{obsid}: 0502091601). In both cases, EPIC pn and MOS were in Small Window
mode (apart from MOS2 in the first observation, which was in Full Frame mode),
which ensures that no significant pile-up is present, as verified with the
{\sc epatplot} tool, in the pn detectors (the only one used for the
analysis). Mrk~704 is by far the brightest source in the field of view.

Data were reduced with \textsc{SAS} 10.0.0, using  
calibration files generated on 2010-6-11. Screening
for intervals of flaring particle background was done consistently
with the choice of extraction radii, in an iterative process based
on the procedure to maximize the signal-to-noise ratio described by
Piconcelli et al. (2005). After this process, the net exposure time was of
about 15 and 68 ks for the 2005 and the 2008 observation, respectively, adopting
extraction radii of 36 and 40 arcsec, and patterns 0 to 4. The background
spectra were extracted from source-free circular regions with a
radius of 50 arcsec.
The same regions were used for the timing analysis. Spectra were binned in order to
oversample the instrumental resolution by at least a factor of 3 and
to have no less than 30 counts in each background-subtracted
spectral channel. The latter requirement allows us to use the
$\chi^2$ statistics as a goodness-of-fit-test.
 RGS source and background spectra were extracted with standard procedures, 
adopting the data reduction pipeline \textsc{rgsproc}, and choosing the NED optical nucleus of 
Mrk~704 as the reference point for the attitude solution.

\section{XMM-$Newton$: Data analysis}

We will use in this paper only data from the EPIC pn and the RGS camera, 
because after the latest release of the EPIC pn redistribution and the RGS
contamination model (May 2010) their cross-calibration is as good as 3\%
(A. Pollock, private communication).

In Fig.~\ref{lc_obs} the 2-10 keV (upper panel), 0.3-2 keV, (middle
panel) and the (2-10 keV)/(0.3-2 keV) hardness ratio (HR) light curves are
shown for the first (left) and the second (right) observations. 
Intra-observation variability in the count rates is apparent in both
cases. While in the first observation no clear variation in the HR is
present, in the second observation an anticorrelation of the hardness ratio
with the flux is evident, indicating that the source gets softer when
brighter. Therefore, while for the first observation we will use, for the
spectral analysis, the time-integrated spectrum only, for the second
observation we will also make a time-resolved spectral analysis.

Spectral analysis has been performed using the XSPEC package, version 12.6.0.
Unless specifically stated, all errors refer to the 90\% confidence level for
one interesting parameter. 

\begin{figure*}
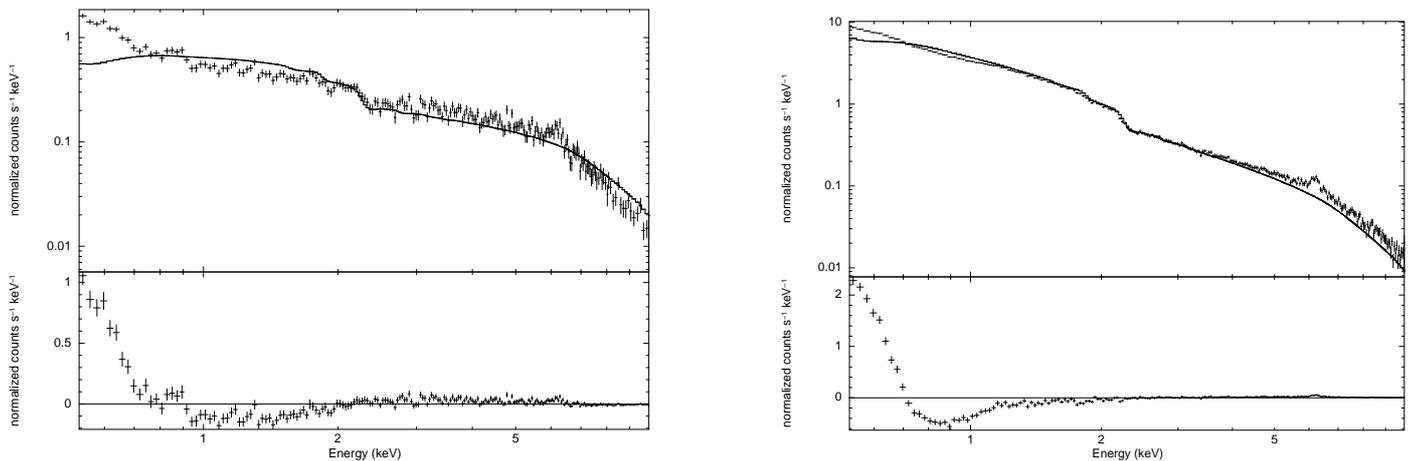

\epsfig{file=badfit_obs1.ps,width=6cm,angle=-90}
\hfill
\epsfig{file=badfit_obs2.ps,width=6cm,angle=-90}
  \caption{The spectrum and residuals for the first (left) and second (Right) XMM-$Newton$ observations when
fitted with a simple power law absorbed by Galactic interstellar matter.}
  \label{badfitobs}
\end{figure*}

Let us start with the time-averaged analysis of both observations. In
Fig.~\ref{badfitobs}  the best fit models and residuals
are shown, assuming a simple power law absorbed by the Galactic column density
in the direction of the source (2.97$\times$10$^{20}$ cm$^{20}$, Kalberla et
al. 2005). The fits are clearly bad, not surprisingly given the extremely
simple model. From the residuals, a soft excess, a broad absorption
trough between about 1 and 2 keV - indicative of obscuration by ionised gas - and an
iron line at 6.4 keV are apparent.

\begin{figure}
\epsfig{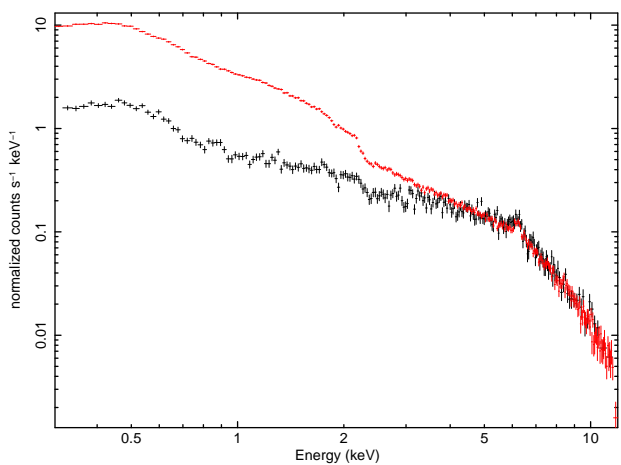}
  \caption{The two XMM-$Newton$ spectra shown together for comparison. The soft
X-ray flux in the first observation is much lower than in the second observation,
while above 5 keV the two spectra are almost indistinguishable.}
  \label{comparison}
\end{figure}

Comparing the spectra of the two observations (Fig.~\ref{comparison}), a huge
difference in the soft part of the spectrum is clear, suggesting a much
stronger absorption in the first observation. Above a few keV, instead, the
two spectra are pretty similar (see also Fig.~\ref{comparison_iron}).

\subsection{The iron line}

We first concentrate on the iron line, and to this end we limit the bandwidth
to 5-10 keV, where the soft excess
and the warm absorber do not have an important impact on the spectral modeling.
More specifically, the
choice of 5 keV as the lower limit of the band is due, on one hand, on the
necessity to have the entire line within the band (which is verified
{\sl a posteriori}), and on the other hand to avoid any effect of the
warm absorber. In fact, a fit with a simple power law gives very different
values - especially for the first observation - of the photon index depending
on the lower limit of the band, stabilizing only above 5 keV. The photon index
steepens increasing the lower limit, a clear indication of absorption at low
energies.

We fitted the 5-10 keV spectrum with a power law, a Compton reflection
component ({\sc pexrav} model in XSPEC, parameter $R$ - the solid angle 
subtended by the reflecting matter to the nuclear source in units of 2$\pi$ -
fixed to 1) and an
iron line, firstly modeled with a narrow Gaussian at 6.4 keV ($\sigma$ fixed
to 0). Results are summarized in Table~\ref{iron}, where
best fit spectral parameters for this and following models are 
reported. In both observations, the
statistical quality of the fit is not very good, and residuals are evident
around the line, suggesting it is resolved. Indeed, leaving $\sigma$ free to
vary, the quality of the fit significantly improves (see Table~\ref{iron}),
with $\sigma$ of 100-150 eV, corresponding to a FWHM velocity 2-3 times larger
than that of the Broad Line Regions (about 5000 km/s, Stirpe 1990). 
Letting the centroid energy of the 
line free to vary, instead, does not improve the quality of the fit.

The most likely broadening mechanism is therefore matter rotation in an accretion
disk. We therefore substituded the Gaussian with the {\sc diskline} model
(valid for Schwarschild metric) with outer radius fixed to 1000 gravitational
radii and the emissivity index to -2.5. The rest-frame line energy is again fixed
to 6.4 keV. The fits are of comparable statistical
quality to those obtained with a broad Gaussian. The moderate width of the
line translates in a quite large inner radius (much larger than the innermost
stable orbit) and in a small inclination angle.

On the other hand, it is possible that the line is actually a blend of a
narrow line (originating e.g. in the torus) and of a relativistic line. We
therefore fitted the spectrum with the combination of the two, fixing for
simplicity the inner radius of the relativistic line to 6 gravitational radii
(the innermost stable orbit for Schwarschild metric) and the inclination angle
to 30 degrees. The fit is again good, and we cannot really distinguish between
the different models on statistical ground. (A fit of similar quality is
obtained if a {\sc laor} model is used for the line, again with the inner
radius fixed to the innermost stable orbit.) We limit ourselves to note that a
relativistic component seems to be required, but whether the line emitting
region is truncated or a narrow component is also present cannot be told.

\begin{figure}
\epsfig{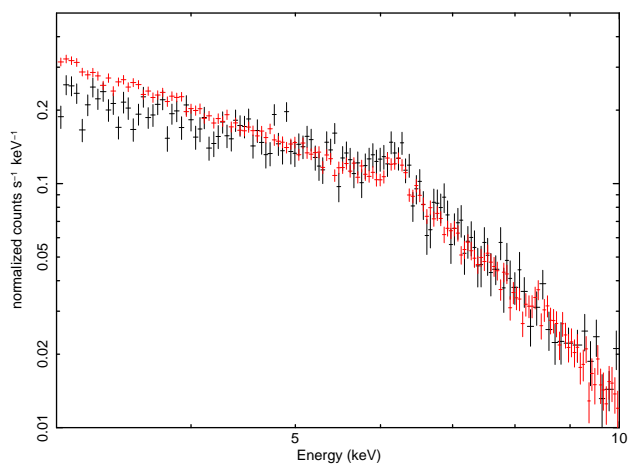}
  \caption{The hard X-ray spectra of the two XM-$Newton$ observations
shown together for comparison.}
  \label{comparison_iron}
\end{figure}

Comparing the two observations, there is marginal evidence for a stronger (and
broader) broad component of the line in the first observation 
(see Fig.~\ref{comparison_iron}). The quality of the data
is, however, not good enough to further explore this possible line
variability.

For both observations, and for all line models, the power law photon index is
around 1.7-1.8.

\begin{table}
  \caption{Iron line results}
  \begin{tabular}{ccc}
    \hline
    &   &  \\
    Parameters &  Obs. 1  &  Obs. 2 \\
    &   &  \\
    \hline
    &   &  \\
    Narrow Gaussian   &   &  \\
    &   &  \\
    F$_{line}$ (10$^{- 5}$ cgs)  &  1.17$^{+ 0.36}_{- 0.37}$  &  0.95$^{+
    0.17}_{- 0.15}$ \\
    EW (eV)  &  85  &  72 \\
    $\chi^2$/d.o.f.  &  77.2/73  &  113.3/93 \\
    &   &  \\
    \hline
    &   &  \\
    Broad Gaussian  &   &  \\
    &   &  \\
    $\sigma$ (eV)  &  143$^{+ 66}_{- 59}$  &  116$^{+ 42}_{- 35}$ \\
    F$_{line}$ (10$^{- 5}$ cgs)  &  2.18$^{+ 0.73}_{- 0.69}$  &  1.51$^{+
    0.32}_{- 0.30}$ \\
    EW (eV)  &  162  &  119\\
    $\chi^2$/d.o.f.  &  66.2/72  &  91.4/92 \\
    &   &  \\
    \hline
    &   &  \\
    Diskline  &   &  \\
    &   &  \\
    $r_{in}$ ($r_g$)  &  52$^{+ 80}_{- 24}$  &  $> 80$ \\
    $\theta$  &  20$^{+ 5}_{- 6}$  &  22$^{+ 20}_{- 7}$ \\
    F$_{line}$ (10$^{- 5}$ cgs)  &  2.24$^{+ 1.01}_{- 0.61}$  &  1.32$^{+
    0.21}_{- 0.25}$ \\
    EW (eV)  &  175  &  113 \\
    $\chi^2$/d.o.f.  &  62.6/71  &  90.7/91 \\
    &   &  \\
    \hline
    &   &  \\
    Diskline + Narrow Gaussian  &   &  \\
    &   &  \\
    F$_{line, disk}$ (10$^{- 5}$ cgs)  &  2.66$^{+ 2.00}_{- 1.28}$  & 
    1.49$^{+ 0.57}_{- 0.58}$ \\
    EW$_{disk}$ (eV)  &  208  &  127 \\
    F$_{line, NG}$ (10$^{- 5}$ cgs)  &  0.68$^{+ 0.46}_{- 0.42}$  &  0.64$^{+
    0.19}_{- 0.20}$ \\
    EW$_{NG}$ (eV)  &  44  &  52 \\
    $\chi^2$/d.o.f.  &  66.7/72  &  96.0/92 \\
    &   &  \\
    \hline
  \end{tabular}{\noindent}
\label{iron}

\end{table}

\begin{table}
  \caption{Warm absorber results}
  \begin{tabular}{ccc}
    \hline
    &   &  \\
    Parameters &  Obs. 1  &  Obs. 2 \\
    &   &  \\
    \hline
    &   &  \\
    1 Zone  &   &  \\
    &   &  \\
    $\Gamma_s$  &  10.0$^{+ 0.0(*)}_{- 0.1}$   &  4.55$^{+ 0.42}_{- 0.25}$ \\
    $\Gamma_h$  &  1.45$^{+ 0.05}_{- 0.04}$  &  1.87$^{+ 0.02}_{- 0.02}$ \\
    N$_H$ (10$^{22}$ cm$^{- 2}$)  &  5.16$^{+ 0.49}_{- 0.34}$  &  0.15$^{+
    0.02}_{- 0.03}$ \\
    $\xi$ (cgs)  &  122$^{+ 9}_{- 8}$  &  39$^{+ 8}_{- 5}$ \\
    E$_{line}$ (keV)  &  0.900$^{+ 0.005}_{- 0.008}$  &  -- \\
    F$_{line}$ (10$^{- 5}$ cgs)  &  4.12$^{+ 0.45}_{- 0.48}$  &  -- \\
    $\chi^2$/d.o.f.  &  295.1/208  &  421.9/230 \\
    &   &  \\
    \hline
    &   &  \\
    2 Zones  &   &  \\
    &   &  \\
    $\Gamma_s$  &  5.4$^{+ 2.4}_{- 2.1}$  &  3.53$^{+ 0.21}_{- 0.14}$ \\
    $\Gamma_h$  &  1.64$^{+ 0.06}_{- 0.07}$  &  1.79$^{+ 0.02}_{- 0.02}$ \\
    N$_{H, 1}$ (10$^{22}$ cm$^{- 2}$)  &  0.14$^{+ 0.08}_{- 0.09}$  & 
    0.085$^{+ 0.007}_{- 0.008}$ \\
    $\xi_1$ (cgs)  &  0.39$^{+ 0.58}_{- 0.18}$  &  5.1$^{+ 0.09}_{- 1.1}$ \\
    N$_{H, 2}$ (10$^{22}$ cm$^{- 2}$)  &  6.85$^{+ 0.57}_{- 0.63}$  & 
    1.63$^{+ 1.15}_{- 0.56}$ \\
    $\xi_2$ (cgs)  &  139$^{+ 13}_{- 13}$  &  1750$^{+ 205}_{- 190}$ \\
    E$_{line}$ (keV)  &  0.904$^{+ 0.005}_{- 0.007}$  &  -- \\
    F$_{line}$ (10$^{- 5}$ cgs)  &  4.57$^{+ 0.49}_{- 0.53}$  &  -- \\
    $\chi^2$/d.o.f.  &  256.4/206  &  279.4/228 \\
    &   &  \\
    \hline
    &   &  \\
    2 Zones, partial covering &   &  \\
    &   &  \\
    $\Gamma_s$  &  6.7$^{+ 3.3(*)}_{-3.2}$   &  2.71$^{+ 0.14}_{- 0.08}$ \\
    $\Gamma_h$  &  1.71$^{+ 0.06}_{- 0.05}$  &  1.18$^{+ 0.33}_{- 0.45}$ \\
    N$_{H, 1}$ (10$^{22}$ cm$^{- 2}$)  &  0.10$^{+ 1.00}_{- 0.05}$  & 
    0.75$^{+ 0.08}_{- 0.06}$ \\
    $\xi_1$ (cgs)  &  0.29$^{+ 1.90}_{- 0.19}$  &  2.24$^{+ 0.38}_{-0.29}$ \\
    Covering Factor (1)  &  $>$0.45  &  0.56$^{+ 0.02}_{- 0.03}$ \\
    N$_{H, 2}$ (10$^{22}$ cm$^{- 2}$)  &  8.8$^{+ 1.8}_{- 1.0}$  & 
    13.3$^{+ 1.9}_{- 2.5}$ \\
    $\xi_2$ (cgs)  &  97$^{+ 15}_{- 11}$  &  122$^{+ 64}_{- 17}$ \\
    Covering Factor (2)  &  0.84$^{+ 0.04}_{- 0.04}$  &  0.38$^{+ 0.11}_{- 0.13}$ \\
    E$_{line}$ (keV)  &  0.895$^{+ 0.011}_{- 0.014}$  &  -- \\
    F$_{line}$ (10$^{- 5}$ cgs)  &  2.50$^{+ 0.78}_{- 0.75}$  &  -- \\
    $\chi^2$/d.o.f.  &  234.3/204  &  230.2/226 \\
    &   &  \\
    \hline
  \end{tabular}{\noindent}
\label{t_zxipcf}
\begin{list}{}{}
\item (*) The asterisk indicates that a parameter is pegged to one of its limits.
\end{list}
\end{table}

\subsection{The warm absorber}

We now study the broad band (0.5-10 keV) spectra of the source. Having in mind
the residuals to a simple power law model (Fig.~\ref{badfitobs}), 
we added to the model already used to study the iron line (adopting
for simplicity a broad Gaussian for the line) a second power law to model the
soft excess, and one or more warm absorber region. The warm absorber was
modeled with the {\sc zxipcf} model, based on the XSTAR photoionization code. 
The main limitations of the model are a fixed
ionizing continuum (a power law with photon index of 2) and fixed element
abundances (equal to solar). The ionization parameter is $\xi = L /
nR^2$. The results are summarized in Table~\ref{t_zxipcf}.
For the first observation, the addition of an emission line 
around 0.9 keV (to be likely identified with the Neon IX He-like triplet) was also
necessary, as also suggested by the RGS analysis. In fact, in the RGS
spectrum of obs 1 a line at 0.903$\pm$0.001 keV (corresponding to the forbidden
line of the triplet) with a flux of
1.69($\pm$0.72)$\times$10$^{- 5}$ ph cm$^{- 2}$ s$^{- 1}$ was found (see below).

\begin{figure*}
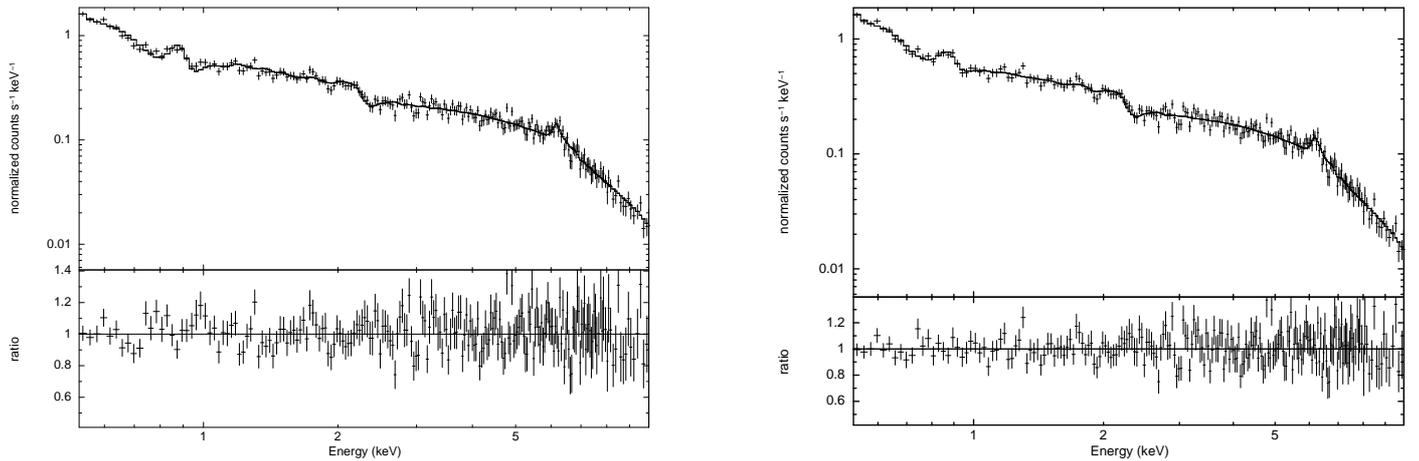

\epsfig{file=bestfit_wa_obs1.ps,width=6cm,angle=-90}
\hfill
\epsfig{file=bestfit_wa_pc_obs1.ps,width=6cm,angle=-90}
  \caption{Obs 1: Best fit model and data/model ratio for the two zones model. Left panel:
full covering absorbers. Right panel: partial covering absorbers. }
  \label{bestfit_obs1}
\end{figure*}

\begin{figure*}
\epsfig{file=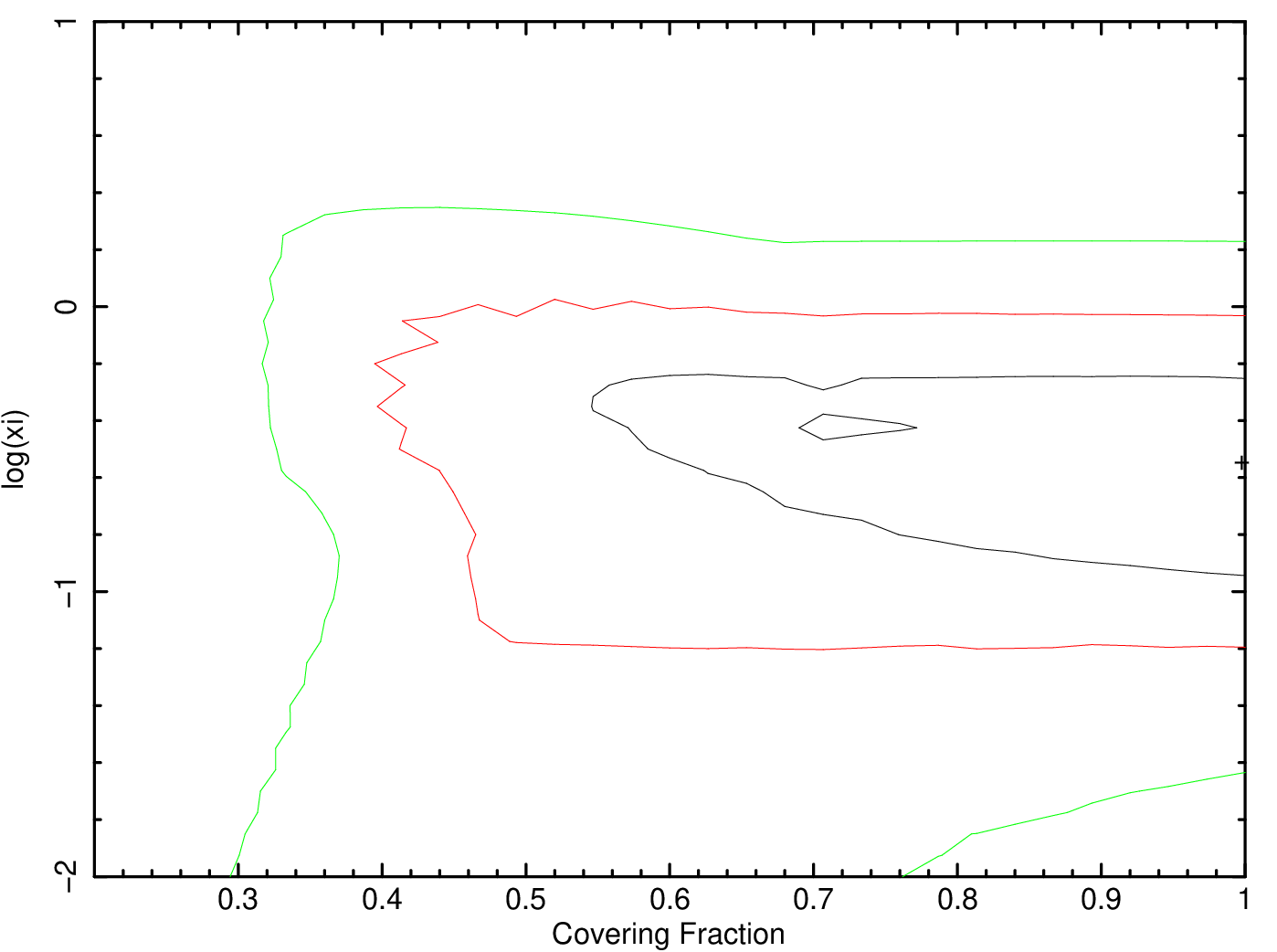,width=6cm,angle=-90}
\hfill
\epsfig{file=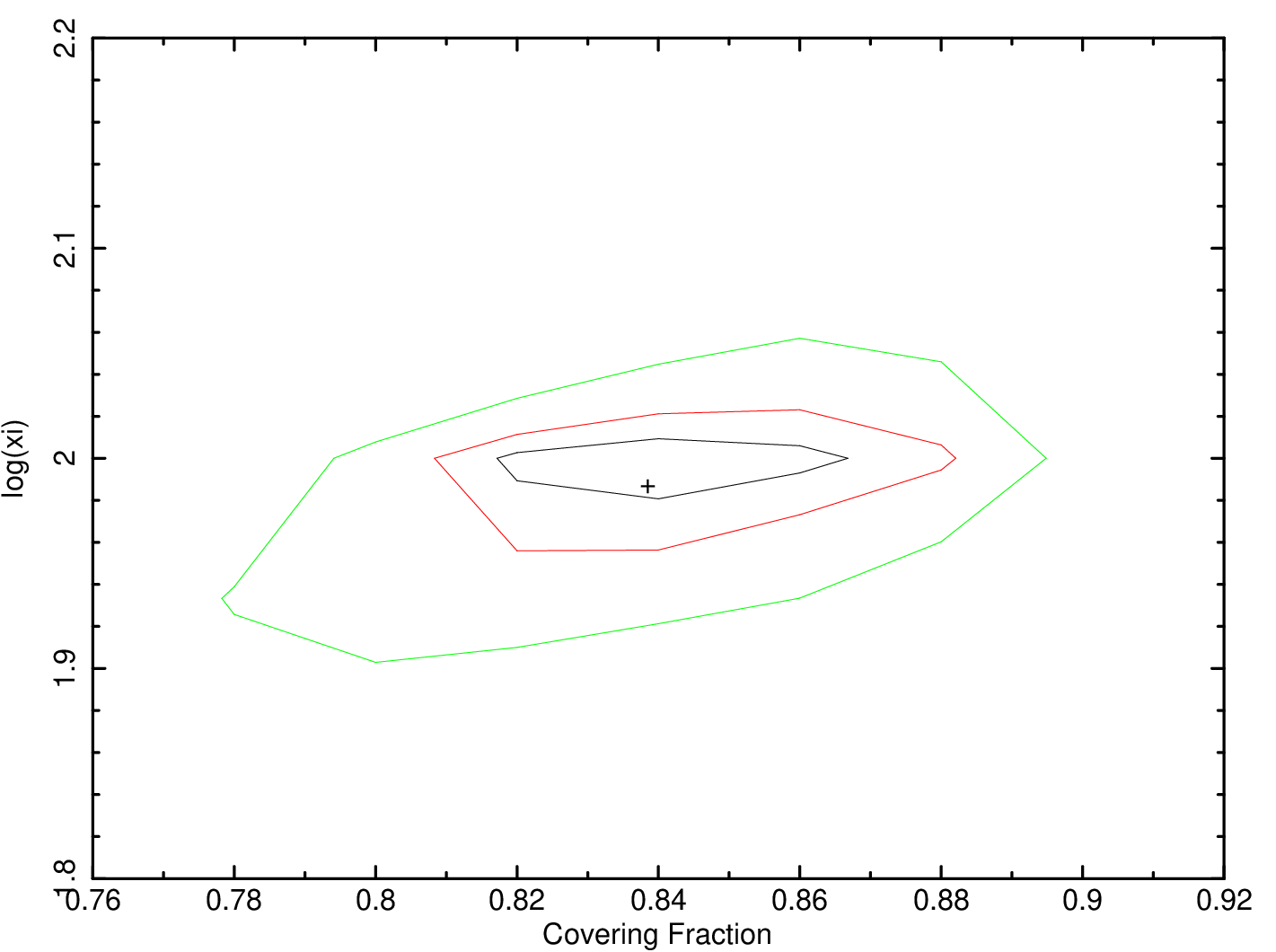,width=6cm,angle=-90}
  \caption{Obs 1: Contour plots (ionization parameters vs covering factors)
for the low (left panel) and high (right panel) ionization absorbers.
The three curves represent the 67\%, 90\% and 99\% confidence levels.}
  \label{contour-1-pc}
\end{figure*}

\begin{table}
  \caption{Velocities of the warm absorber regions}
  \begin{tabular}{ccc}
    \hline
    &   &  \\
    Models &  Obs. 1  &  Obs. 2 \\
    &   &  \\
    \hline
    &   &  \\
    2 Zones  &   &  \\
    &   &  \\
    $v_{low}$ & -3900$^{+3300}_{-6500}$ km/s & -190$\pm$210 km/s \\
    $v_{high}$ & +2650$^{+9300}_{-870}$ km/s & -100$\pm$250 km/s\\
    &   &  \\
    \hline
    &   &  \\
    2 Zones, partial covering &   &  \\
    &   &  \\
    $v_{low}$ & --  & -580$\pm$340\\
    $v_{high}$ & +3000$^{+1500}_{-1800}$ km/s & +140$\pm$300 km/s \\
    &   &  \\
    \hline
  \end{tabular}{\noindent}
\label{t_vel}
\end{table}

\subsubsection{The first observation} 


Let us first discuss the first observation,
where the absorption is much stronger. A single ionization zone is 
insufficient, the addition of a second zone improving significantly, even if not dramatically, 
the statistical quality of the fit. Moreover, in the single zone model
$\Gamma_s$ (the photon index of the soft power law) is unrealistically steep, while
$\Gamma_h$ (the photon index of the hard power law) 
is much flatter than the value found in the hard band.

The addition of the second zone improves the quality of the fit, which however
remains not fully satisfactory. Visual inspection of the data/model ratio
(see Fig.~\ref{bestfit_obs1}), however, shows that this is due mostly to noise rather than to
unfitted features, even if some wiggles are present around the Neon line. 
Indeed, an improvement ($\Delta \chi^2$=-16) of the
fit is obtained by letting free to vary the width of the line, when a value of
48 eV is found. This may suggest a blend with other lines (mostly iron L), too
faint to be individually detected by the RGS. However, it is also possible that the
broadening of the line is an artifact to correct for a too strong, in the
model, iron UTA (this may occur e.g. if the iron abundance is lower than solar). 
Unfortunately, with the present data we cannot go much further in this analysis.

No further improvement is found by adding a third absorption region.

We tried to fit the warm absorber also with ``home-made'' tables built using
the {\sc cloudy} photoionization code. Results are similar 
and are therefore not reported here.

As a consistency check, we then fitted the RGS data with the best fit model
(all parameters fixed) described above. Spectra were rebinned in order to have
at least 10 counts per bin\footnote{As a check that the rebinning was not altering
the fit results, we fitted the RGS spectra also with no rebinning and the Cash
quality-of-the-fit statistics, finding no significantly different results}.
The fit is reasonable ($\chi^2$/d.o.f.=333.4/278),
with several, but not individually significant features apparent in the residuals,
apart from the Ne IX line already included in the spectrum. 
An improvement  ($\chi^2$/d.o.f.=296.2/269)
is found letting the model parameters free to vary. The best fit parameters, however, 
are consistent within the errors with those found in the EPIC-pn analysis, and the abovementioned
residuals are still there. 

We then let the velocities of the absorbing zones free to vary, to account for
possible inflows/outflows. A better fit is found  ($\chi^2$/d.o.f.=281.2/267), with an 
inflow velocity of 3900$^{+6500}_{-3300}$ km/s for the low ionization absorber and an
outflow velocity of 2650$^{+9300}_{-870}$ km/s for the high ionization absorber (see 
Table~\ref{t_vel} where ``low'' and
``high'' refers to the low and high ionization components, respectively, and negative
velocities indicate inflows).

Back to the EPIC-pn analysis, a slight improvement
($\Delta \chi^2$=-8) is also found substituting the soft power law with a 
disk thermal component ({\sc diskbb} model in XSPEC). The temperature is around
160 eV, but the normalization value is unphysical, corresponding to about 4$\times$10$^5$
km for the inner radius (the Schwarzschild radius for the 100 million solar masses 
black hole estimated for Mrk~704, Wang et al. 2009, is about one thousand times larger).

A slightly worse fit  ($\Delta \chi^2$=14) is instead found if an ionized reflection model
({\sc reflionx}, Ross \& Fabian 2005) is adopted (with or without relativistic
blurring) for the soft X-ray component. 
The power law index is pegged to the lower possible value, 1.4.


A moderate improvement ($\Delta \chi^2$=-22) is found 
letting the absorbers be partial. Best fit covering factors of 1 (but loosely constrained)
and 0.85 are found for the colder and warmer ionization zones, 
respectively (see Table~\ref{t_zxipcf}).
All other parameters are similar to those found in the full absorbers fit. 
The best fit model and data/model ratio can be found in Fig.~\ref{bestfit_obs1}.
In Fig.~\ref{contour-1-pc} the contour plots (ionization parameter vs. covering factor)
are shown. For the low ionization absorber, a full covering is always acceptable, 
while at the 99\% confidence level a very low ionization (almost neutral) partial absorber is
possible. The two parameters are much better determined for the high ionization absorber, 
with no significant correlation between them. 

\begin{figure}
\epsfig{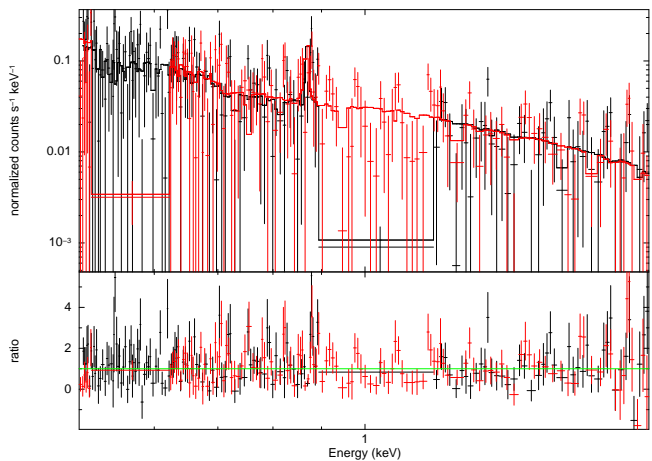}
  \caption{RGS best fit spectra and data/model ratio for the
first observation. Fit with partial covering absorbers.}
  \label{rgsobs1}
\end{figure}

Regarding the RGS, a check similar to that described above for the full covering scenario
was also performed in the partial covering case.
With all parameters fixed, the fit is worse than for the full absorbers  ($\chi^2$/d.o.f.=356.5/278).
A slightly better fit is instead found letting the
parameters free to vary ($\chi^2$/d.o.f.=283.7/267), again with the best fit parameters 
consistent within the errors with those found in the EPIC-pn analysis. The feature-like residuals
are still apparent (see Fig.~\ref{rgsobs1}). 
Letting the velocities of the absorbers free to vary, we found a marginal improvement
of the quality of the fit ($\chi^2$/d.o.f.=279.2/265),  with an
outflow velocity of 3000$^{+1500}_{-1800}$ km/s for the high ionization absorber, while
the velocity of the low ionization absorber is basically undetermined (Table~\ref{t_vel}).

The value of $\Gamma_h$ is harder than found in the iron line
analysis. This, together with the relatively high $\chi^2$, may indicate that
the warm absorber model we are adopting is not fully adequate (e.g.,
the absorption regions may be more structured that single ionization zones,
or metal abundances be different than solar). 

For simplicity, in the broad band fitting the iron line is modeled with a broad
Gaussian. No significant changes in the line parameters are found with 
respect to the hard X-ray band fits described in the previous section.

The 0.5-2 (2-10) keV observed flux is 1.53(10.1)$\times$10$^{- 12}$ erg 
cm$^{-2}$ s$^{-1}$, corresponding to a 0.5-2 (2-10) keV luminosity of
1.17(2.34)$\times$10$^{43}$ erg s$^{-1}$, after correction for absorption.

\subsubsection{The second observation}


Due to the much weaker absorption and the absence of the Neon emission
line, and despite the better statistics, the analysis of the second
observation resulted to be easier. Also in this case, however, 
the addition of a second zone is definitely required.

\begin{figure*}
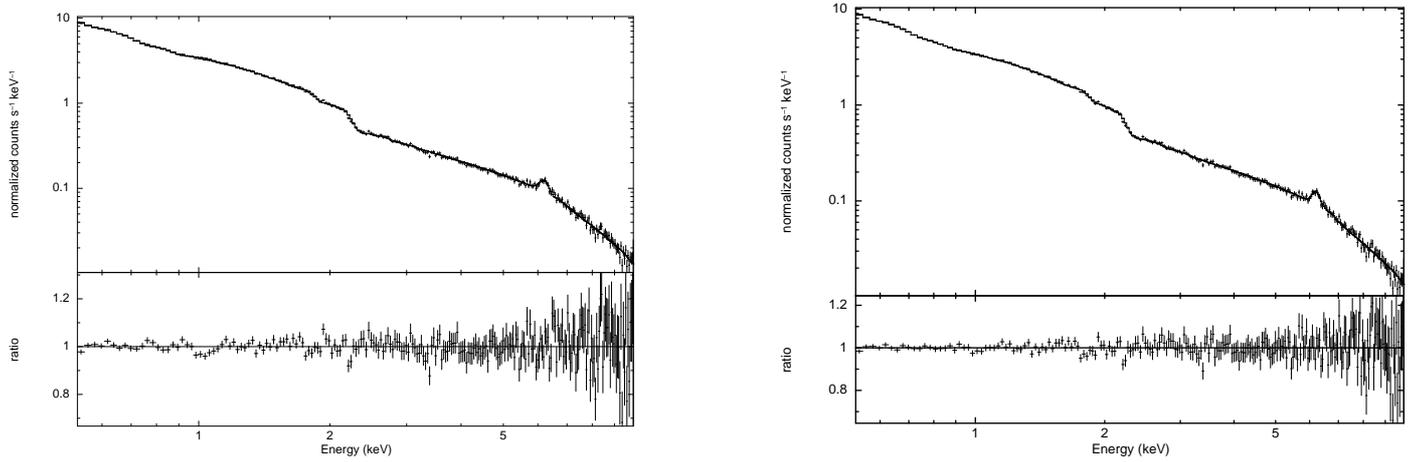

\epsfig{file=bestfit_wa_obs2.ps,width=6cm,angle=-90}
\hfill
\epsfig{file=bestfit_wa_pc_obs2.ps,width=6cm,angle=-90}
  \caption{Obs 2: Best fit model and data/model ratio for the two zones model. Left paneL:
full covering absorbers. Right panel: partial covering absorbers.}
  \label{bestfit_obs2}
\end{figure*}

Inspection of the data/model ratio (see Fig~\ref{bestfit_obs2}) shows
however that some features are still remaining, even if that 
at about 1.8-2 keV is very likely of instrumental origin.
Particularly prominent is the feature around 1 keV, again suggesting a
possible problem in the fitting of iron transitions. The addition of a third
absorbing zone does not cure this problem.

Also for this observation we find similar results with the {\sc cloudy}-based, home-made
tables. A significantly worse fit is found, instead, substituting the soft power law
either with a disk thermal component or with an ionized reflection model.

We then fitted the RGS data with the best fit model
(all parameters fixed) above described. Given the much better statistics with respect to the
first observation, spectra were
rebinned to have at least 50 counts per bin\footnote{Also in this case, fits to the unbinned
spectra provide substantially the same results}. The fit is reasonable ($\chi^2$/d.o.f.=856.8/679). 
An improvement  ($\chi^2$/d.o.f.=814.4/672)
is found letting the model parameters free to vary. The best fit parameters, however, 
are consistent within the errors with those found in the EPIC-pn analysis. Inspection of the residuals
indicates the possible presence of features, but the only significant one
(at the 99\% confidence level)  
is an emission line with a centroid energy of 1.114$\pm$0.002 keV - to be possibly identified with 
Fe XVII L lines - with a flux of 1.80($\pm$0.75)$\times10^{-5}$ ph/cm$^{2}$/s$^{-1}$. 
However, no such line is found in the EPIC-pn data (upper limit of  3.6$\times10^{-6}$ ph/cm$^{2}$/s$^{-1}$).
Finally, no significant inflow/outflow velocity is found, with upper limits of a few hundreds km/s
(Table~\ref{t_vel}).

\begin{figure*}
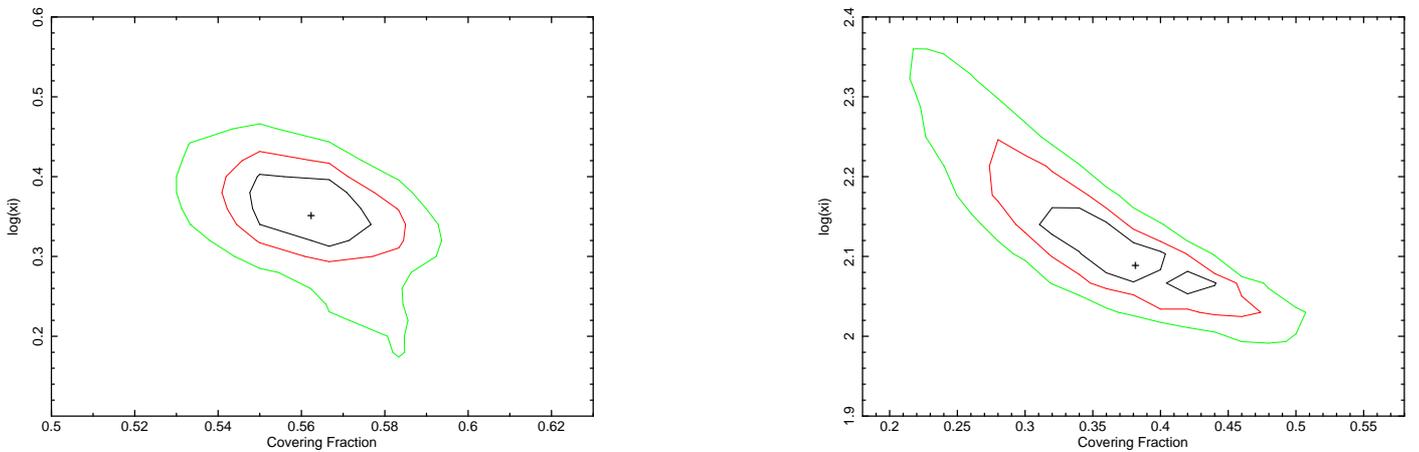

\epsfig{file=contour_2_PC_1.ps,width=6cm,angle=-90}
\hfill
\epsfig{file=contour_2_PC_2.ps,width=6cm,angle=-90}
  \caption{Obs 1: Contour plots (ionization parameters vs covering factors)
for the low (left panel) and high (right panel) ionization absorbers.
The three curves represent the 67\%, 90\% and 99\% confidence levels.}
  \label{contour-2-pc}
\end{figure*}

A significant improvement ($\Delta \chi^2$=-49) is found 
letting the absorbers be partial. Best fit covering factors of 0.56 (but loosely constrained)
and 0.38 are found for the colder and warmer ionization zones, 
respectively (see Table~\ref{t_zxipcf}).
Both absorbers are now much thicker and slightly less ionized. The power law indices are much harder;
in particular, the $\Gamma_h$ is now 1.18 (even if loosely constrained), a value unusually hard for 
Seyfert galaxies. However, it must be noted that fixing  $\Gamma_h$=1.8 a good fit is still obtained
($\chi^2$=243.0/227), with the parameters of the less ionized absorber almost unchanged while the 
more ionized absorber gets still more ionized and thicker 
($N_{H,2}\sim$4$\times$10$^{23}$ cm$^{-2}$, $\xi_2\sim$230, covering factor of 0.27).
The best fit model and data/model ratio can be found in 
Fig.~\ref{bestfit_obs2}.
In Fig.~\ref{contour-2-pc}, the contour plots (ionization parameter vs. covering factor)
are shown. For the low ionization absorber the two parameters are almost uncorrelated, while
for the high ionization absorber they are anticorrelated, indicating a certain
degree of degeneracy in the model.

\begin{figure}
\epsfig{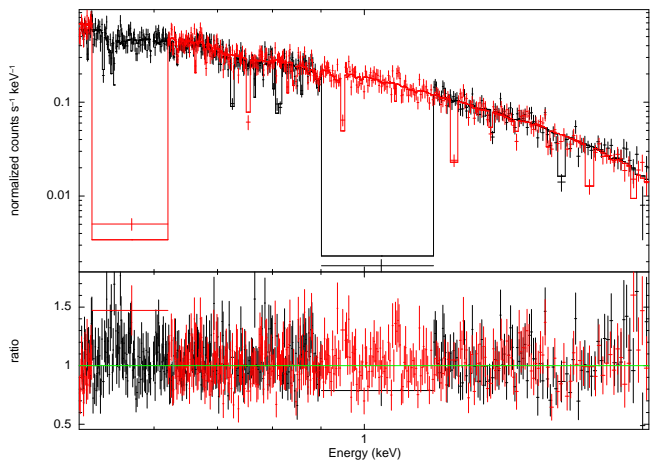}
  \caption{RGS best fit spectra and data/model ratio for the
second observation. Fit with partial covering absorbers.}
  \label{rgsobs2}
\end{figure}

The usual check with the RGS is then performed. Fits are better than those obtained
with the full absorbers, reflecting the same improvement found in the EPIC-pn spectrum.
With all parameters fixed, a $\chi^2$/d.o.f.=821.6/679 is found, while letting
the  parameters free to vary, a  $\chi^2$/d.o.f.=752.0/670 is found. 
The best fit parameters are
consistent within the errors with those found in the EPIC-pn analysis. The spectrum and data/model
ratio can be seen in Fig.~\ref{rgsobs2}. 
Finally, no significant inflow/outflow velocity is found for the high ionization absorber (upper limits
of a few hundreds km/s), 
while for the low ionization absorber a marginally significant inflow velocity of 580$\pm$340
km/s is obtained (Table~\ref{t_vel}). 

For the second observation, the 0.5-2 (2-10) keV observed flux is
7.62(11.2)$\times$10$^{- 12}$ erg cm$^{- 2}$ s$^{-1}$, corresponding to a
0.5-2 (2-10) keV luminosity of 2.34(2.26)$\times$10$^{43}$ erg s$^{-1}$,
after correction for absorption. 

During the revision of the present work, we became aware of a paper by Laha et al. 
(2011) dealing with the analysis of the second XMM-$Newton$ observation
of Mrk~704. The results of their analysis are qualitatively similar to ours, the quantitative
differences probably mostly related to the somewhat different spectral model 
adopted\footnote{The main difference in the modeling
is that they included a partial, neutral intrinsic absorber, while keeping full coverage for 
the ionized absorbers. If we did the same for the two full absorbers model, a slightly
better fit is found, which however is significantly worse than the one with two partial 
ionized absorbers.
If we did the same for the first observation, a fit very similar to that with 
two partial ionized absorbers
is found, not surprisingly as the lower ionization zone is almost neutral in that observation
(see Fig.~\ref{contour-1-pc}).}. 
The most important qualitative difference is that Laha et al. found the two warm absorbers
both outflowing, while in our analysis no significant inflow/outflow is found, with only marginal
evidence for an inflow of the low ionization absorber in the partial covering scenario.

\subsubsection{Comparison between the two observations} 

A comparison between the results of the fits to the two spectra of the two
observations show, as obvious from 
Fig.~\ref{comparison}, stronger overall 
absorption in the first observation. Assuming full covering absorbing
zones, for both of them the
column density decreases (by a factor about 2 for the colder absorber, a
factor about 4 for the warmer absorber) and the ionization parameter increases
from the first to the second observation.
If the absorbers are allowed to partially cover the source, the colder absorber
results thicker, warmer and with a lower covering factor in the second observation,
while for the warmer zone the most significant difference is the covering factor,
again much lower in the second observation (the column density and the ionization 
parameter both slightly increase).

The detection of a Neon emission line in the first observation only cannot be entirely
ascribed to the stronger absorption (which would make by contrast a constant line
more visible). In fact, simulations show that a line with that flux should have
been detected also in the second observation (even if the upper limit to such a line,
about 10$^{-5}$, is marginally consistent with the value found in the RGS in the
first observation). Therefore, an intrinsic variability of the line intensity is
required, possibly related to a different covering factor of the absorbing/reflecting
clouds.

So, the question is: what drove a so large variation of the warm
absorber over three years? The hard 2-10 X-ray luminosity stayed
approximately the same, while the soft 0.5-2 keV luminosity was higher by a 
factor ~2 in the second observation (even if the reconstructed values, given
the large absorption, are rather uncertain), which may - in part at least -
explain the higher ionization parameters. However, the different values in the
two observations of the column densities and of the Neon line suggest that
also variations in the general properties (size, distance, covering fraction, etc.)
of the circumnuclear medium have occurred.

\subsubsection{Time dependent analysis of the second observation}

\begin{figure}
\epsfig{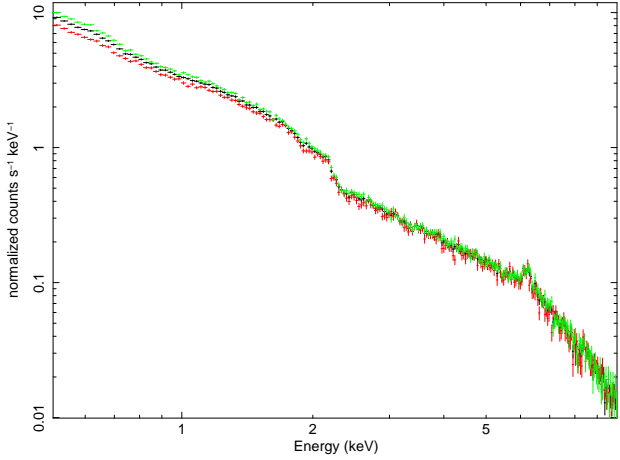}
  \caption{The spectra from the first and second part of the second 
XMM-$Newton$ observation, shown together for comparison.}
  \label{timeres}
\end{figure}

Due to its intra-observation spectral variability (see Fig.~\ref{lc_obs}), we
divided the second observation in two parts, corresponding to the first 40
ksec (obsA here-in-after) and to the remaining part of the observation (obsB),
respectively. The three spectra (obsA, obsB and total) are shown in
Fig~\ref{timeres}. A reasonable fit ($\chi^2$/d.o.f. of 266.2/228 and 274.5/235 for
obsA and obsB, respectively) is obtained by imposing all spectral parameters
to be fixed to those found in the total spectrum (two absorbing zones model), 
leaving as free parameters
only the normalizations of the two power laws. The observed spectral
variability is then explained mainly in terms of a variation of the normalization of
the soft power law. Other choices, like leaving instead as a free parameter
the column density or the ionization parameter or the covering factor
of the warm absorbers do not provide acceptable fits.

Leaving all main parameters free to vary, better fits are of course obtained
($\chi^2$/d.o.f=253.5/221 and 265.2/228 for obsA and obsB, respectively) but the
parameter which suffered the largest variation is 
again the normalization of the soft power law, confirming
the previous analysis.

\section{The Swift and $ASCA$ observations}

\begin{table*}
  \caption{Swift/XRT results}
  \begin{tabular}{ccccc}
    \hline
    &   &  & & \\
    Parameters &  Obs. 1  &  Obs. 2 & Obs. 3 & Obs 5\\
    &   &  &  & \\
    \hline
    &   &  &  & \\
Date & 2006-01-06 & 2006-06-14 & 2006-09-28 & 2007-01-21  \\
XRT exp. time (s) & 671 & 2258 &  5602 & 1750 \\
    $\Gamma$  &  2 (fixed) & 2.00$^{+0.21}_{-0.16}$  &  1.88$^{+ 0.11}_{-0.12}$ & 2 (fixed) \\
    N$_H$ (10$^{22}$ cm$^{- 2}$)  &  12$^{+8}_{-8}$ & 0.85$^{+ 2.20}_{- 0.57}$  &  
1.58$^{+1.35}_{-1.06}$ &  $<$1.1  \\
    $\xi$ (cgs)  & 95$^{+27}_{-81}$ & 73$^{+325}_{-53}$  &  123$^{+172}_{-97}$ &   $<450$ \\
    Flux (2-10 keV) (10$^{-11}$ cgs) & 0.64 & 1.14 & 0.88 & 0.87  \\
    $\chi^2$/d.o.f.  &  0.31/5 & 38.4/36  & 49.1/54 & 19.3/22   \\
    &   &  & \\
    \hline
  \end{tabular}{\noindent}
\label{swift}
\end{table*}

\begin{figure}
\epsfig{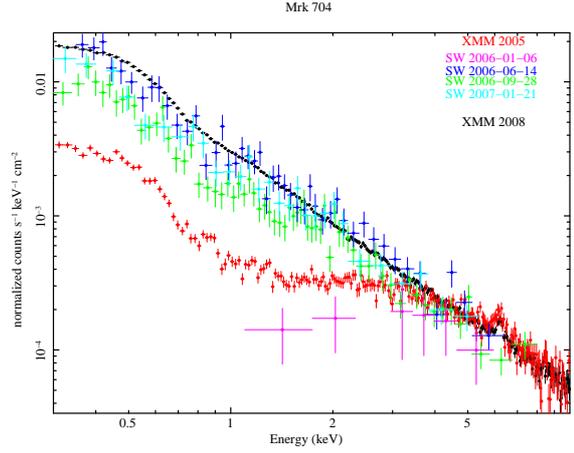}
  \caption{The spectra of all XMM-$Newton$ and Swift observations
shown together for comparison.}
  \label{swiftspectra}
\end{figure}

The source was observed by Swift five times. In four occasions, it was also
in the XRT field of view (see Table~\ref{swift}), while in the fourth observation
it was outside. The four XRT spectra, along with the two XMM ones, 
are shown in Fig.~\ref{swiftspectra}. 

From the figure, spectral and flux variability is apparent between Swift
observations (even if far less dramatic than that between the two, 3-years apart
XMM-$Newton$ observations), implying a variability time scale of a few months or less. In 
Table~\ref{swift} the results of the spectral fits are summarized. Due to the modest
quality of the spectra, a good fit is already obtained with a single power law absorbed
by a single zone, fully covering absorber (no improvement is found letting the covering of the
absorber be partial). For the lesser quality first and last observations, 
we fixed the power law index
to 2, a value similar to those found in the second and third observations. The quality of the
spectra is insufficient to establish if the variability is due to a variation of the properties
of the warm absorber, apart from the first observation, when a much larger absorption is
apparent.

\begin{figure}
\epsfig{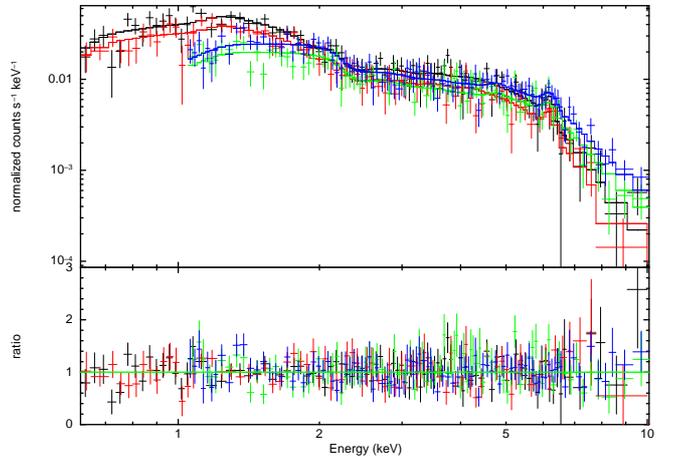}
  \caption{Best fit model and data/model ratio for the $ASCA$ spectrum. See
text for detail.}
  \label{ascabestfit}
\end{figure}

Mrk 704 was also observed by $ASCA$ on May 12, 1998, for about 35 ks exposure time.
A good fit ($\chi^2$/d.o.f=348.9/355) is already obtained with a single power law
absorbed by a single zone, fully covering absorber (see Fig.~\ref{ascabestfit}). 
The source was in a lower flux state (2-10 keV flux of 
about 5$\times$10$^{-12}$ erg cm$^{-2}$ s$^{-1}$. The absorber has a column density
of 5.6($^{+3.5}_{-1.4}$)$\times$10$^{23}$ cm$^{-2}$, larger than in any other observation,
and an ionization parameter of 1430($^{+270}_{-130}$) (cgs).

\section{Discussion and conclusions}

The Seyfert 1.2 galaxy Mrk~704 was observed twice by XMM-Newton, about three years
apart (first observation in October 2005, the second in November 2008).  While the hard
($>$5 keV) spectrum was almost constant between the two observations (with a possible
broadening/strenghtening of the iron line in the first observation), the soft part
dramatically changed, showing much heavier absorption in the first observation.
A strong Neon IX line is also clearly present in the first observation, while it is not
detected in the second one. The very similar hard X-ray spectrum in the two
observations (the behaviour 
of the intrinsic soft X-ray spectrum is more uncertain because of the strong absorption),
together with the change in the column density, suggest that 
the variations are due to a change in the properties
of the absorbing clouds. Dramatic changes of the column density and/or the ionization state
of the absorber have already been observed in other sources, the most spectacular being    
NGC 4151 (e.g. Schurch \& Warwick 2002, de Rosa et al. 2007), NGC 7582 (Bianchi et al. 2009) 
and NGC 1365 (Risaliti et al. 2005, 2009). In the last two sources, in particular, 
large variations of the column density of the (cold) absorber occur on time scales as short as 
less than a day. Here we observe also a strong
variation of the ionization state of the absorbers. As the primary continuum (at least the
hard one) stayed almost constant, this indicates a variation of the location and/or of the density
of the clouds.

It is interesting to note that Mrk~704 is a 
``polar-scattering'' Seyfert 1, i.e. a source with optical polarization 
aligned perpendicularly to the radio source axis, as usually found in Seyfert 2s 
(Smith et al., 2004). Smith et al. suggest that in these sources the nucleus is seen
through the edge of the torus. 

We fitted the spectra with two absorbing regions, either fully or partially covering the
primary emission. The improvement in the quality of the fit when letting  the aborbers be partial
is only moderate for the first observation, but more significant in the second one. 
As the results are somewhat different in the two cases, let us discuss them separately.

With the full covering absorbers, both absorbing zones are found more ionized and with 
a lower column density in the second observation (even if the difference in
the column density for the low ionization zone is only marginal). The RGS analysis of the first observation
suggests a possible (but only marginally significant) inflow
for the low ionization, low column density absorber, and an outflow velocity 
(very poorly determined, with 90\% confidence
level range between 1800 and 12000 km/s) for the high ionization, high column density absorber. 
An ionized and unstable torus wind, as suggested by Smith et al., may indeed provide 
an explanation for the latter zone. 
No significant inflow/outflow is instead found, for both absorbers, in the second observation.

The partial covering scenario, even if providing better fits, is more 
demanding geometrically,
requiring a size of the obscuring clouds of the same order of
the size of the emitting region. This naturally points to a much closer location
of the obscuring clouds, as e.g. due to a radiatively-driven accretion disk wind
(e.g. Proga 2003). Interestingly, the variability behaviour of Mrk~704 closely
resemble that of mini-BAL QSOs (Giustini et al. 2010), strengthening the disk wind scenario,
at least for the more ionized absorber (the colder absorber may be composed of orbiting clouds
as those found in NGC~1365, Risaliti et al. 2009). Interestingly, a possible outflow 
has been detected in the first observation (but not in the second, despite very similar 
column densities and ionization parameters).

The alternative hypothesis, that the partial covering is mimicking
the presence of a scattering component originating outside the absorbing region, is ruled out
by the short term variability of the X-ray emission. 

Mrk~704 was also observed four times by Swift/XRT between January 2006 and January
2007. Spectral and flux variability is apparent between the observations, but the quality
of the spectra is not sufficient to establish the nature of the variations.

The iron K$\alpha$ line, studied with XMM, is broad, much broader than the optical 
broad lines but narrower than expected if the entire line would be originated in 
the innermost accretion disk. However, if a narrow
component (which seems to be almost ubiquitous in Seyfert galaxies) is added, then
the remaining broad component is consistent with emission down to the innermost stable
orbit, even if the quality of the data in not good enough to constrain the spin of the
black hole.

\section*{Acknowledgements}

We thank the anonymous referee for her/his suggestions which
helped improving the clarity of the paper.
We thank all the members of the $FERO$ collaboration for useful discussions.
GM, SB and EP acknowledge financial support from ASI under grant I/088/06/0
and I/090/10/0/.
POP acknowledges financial support from CNES and French GDR PCHE.
GP acknowledges support from an EU Marie Curie Intra-European
Fellowship under contract no. FP7-PEOPLE-2009-IEF-254279.

\end{document}